\documentclass[12pt]{article}
\begin{document}

\setlength{\unitlength}{1mm}

 \title{A theoretical construction of wormhole supported by Phantom Energy }
 \author{\Large $F.Rahaman^*$ , $M. Kalam^{**}$, $ M. Sarker ^*$ and $ K. Gayen ^*$ }
\date{}
 \maketitle
 \begin{abstract}
                  A new solution has been presented for the
                  spherically symmetric space time describing
                  wormholes with Phantom Energy. The model
                  suggests that the existence of wormhole is
                  supported by arbitrarily small quantity of
                  Phantom Energy.

  \end{abstract}



 \bigskip
 \medskip
  \footnotetext{ Pacs Nos :  04.20 Gz,04.50 + h, 04.20 Jb   \\

                              $*$Dept.of Mathematics, Jadavpur University, Kolkata-700 032, India\\
                                  E-Mail:farook\_rahaman@yahoo.com

                             $**$Dept. of Phys. , Netaji Nagar College for Women ,
                                          Regent Estate,
                                          Kolkata-700092, India
                              }

    \mbox{} \hspace{.2in}
Wormholes are geometrical structures connecting two distinct
space times. It is very interesting object in Modern science as
because if traversable wormhole exists , the time machine can be
constructed [1]. Wormhole is the solution of Einstein equation
and shared by the violation of null energy condition . The matter
that characterized the above stress energy tensor is known as
exotic matter. There has been a fairly large amount of
discussions [2]on wormholes beginning with the work of Morris and
Throne[3]. Several authors have discussed wormholes in scalar
tensor theory of gravity in which scalar field may play the role
of exotic matter [4]. Recently, Researcher are interested to know
how much exotic matter is needed to get a traversable wormhole
[5-8].  Recent Astrophysical observations indicate that the
Universe at present is accelerating . There are different ways of
evading these unexpected behavior . Most of these attempts focus
on Alternative gravity theories or the supposition of existence
of a hypothetical dark energy with a positive energy density and
a negative pressure [9] . The matter with the property, energy
density, $\rho > 0 $ but pressure $ p < -\rho < 0 $ is known as
Phantom Energy . The Phantom Energy violates the null energy
condition what is needed to support traversable wormhole. Several
authors have recently discussed the physical properties and
characteristics of traversable wormholes by taking Phantom Energy
as source [10-12]. \\
Recently, Zaslavskii [13] have found a particular exact solution
describing wormhole with a linear equation of state , $ p_r <
-\rho < 0 ,   p_{tr} > 0 $ ( $ p_r$ and $p_{tr}$ are radial and
transverse pressures respectively ) . In this article, we try to
carry out the most general study of the existence of possible
static spherical symmetric solution describing wormholes with
Phantom Energy as source . \\
A static spherically symmetric Lorentzian wormhole can be
described by a manifold $ R^2 X S^2 $ endowed with the general
metric in Schwarzschild co-ordinates $( t,r,\theta,\phi )$ as
\begin{equation}
                ds^2 = - e^{2f(r)} dt^2 + \frac{1}{[1 - \frac{b(r)}{r}]}dr^2+r^2 d\Omega_2^2
            \label{Eq1}
          \end{equation}
where,   $ r     \epsilon   (-\infty , +\infty) $ . \\
To represent a wormhole , one must impose the following
conditions on the  metric (1) as [1] :  \\
1) The redshift function, $f(r)$ must be finite for all values of
$r$ . This means no horizon exists in the space time . \\
2) The shape function, $b(r)$ must obey the following conditions
at the throat $ r = r_0 $ : \linebreak $b(r_0) = r_0$ and
$b^\prime(r_0) < 1 $ [these are known as Flare-out conditions].\\
3) $\frac{b(r)}{r} < 1 $ for $ r >r_0 $ i.e. out of throat .\\
4)The space time is asymptotically flat i.e.$\frac{b(r)}{r}
\rightarrow 0 $ as $ \mid r \mid \rightarrow \infty $.\\

Using the Einstein field equations
 $G_{\mu\nu} = 8\pi T_{\mu\nu} $, in orthonormal reference frame
 ( with $ c = G = 1 $ ) , we obtain the following stress energy
scenario,
\begin{equation}
                \rho(r) =\frac{b^\prime}{8\pi r^2}
                \label{Eq2}
          \end{equation}

\begin{equation}
                p(r) =\frac{1}{8\pi} [ - \frac{b}{r^3} + 2\frac{f^\prime}{r} ( 1 -
                \frac{b}{r})]
                \label{Eq3}
          \end{equation}

\begin{equation}
                p_{tr}(r) =\frac{1}{8\pi}( 1 -
                \frac{b}{r}) [ f^{\prime\prime} - \frac{(b^\prime r - b )}{2r(r-b)}f^\prime
              + {f^\prime}^2 + \frac{f^\prime}{ r } - \frac{(b^\prime r -
              b)
              }{2r^2(r-b)}]
                \label{Eq4}
          \end{equation}
where $\rho(r) $ is the energy density, $p(r)$ is the radial
pressure and $p_{tr}(r)$ is the transverse pressure. \\

Using the conservation of stress energy tensor $
T^{\mu\nu}_{;\nu} = 0 $, one can obtain the following equation
\begin{equation}
                p^\prime  + f^\prime \rho + ( f^\prime +  \frac{2}{r})p -
                \frac{2}{r}p_{tr} = 0
                \label{Eq5}
          \end{equation}

 From now on , we assume that our source is characterized by the
 Phantom Energy with equation of state that contains a radial
 pressure
 \begin{equation}
               p =  - k \rho
                \label{Eq6}
          \end{equation}
we suppose also that pressures are isotropic and
\begin{equation}
               p_{tr} = k  \rho
                \label{Eq7}
          \end{equation}
[ here k is either$ >1$ or $<-1$ ]

Since only two equations of the system (2) - (4) are independent
, it is convenient to represent them as follows:
 \begin{equation}
               {b^\prime}= {8\pi r^2}\rho(r)
                \label{Eq8}
          \end{equation}
\begin{equation}
                {f^\prime}=\frac{(8\pi p r^3 + b )}{2r(r-b)}
                \label{Eq9}
          \end{equation}
From (5) by using (6) and (7), one can obtain
\begin{equation}
                \rho(r)e^{(1-\frac{1}{k})f} =\frac{\rho_0}{r^4}
                \label{Eq10}
          \end{equation}
where $ \rho_{0} $ is an integration constant.\\

Taking into account equations (8),(9) and (10), we have the
following equation containing 'b' as
\begin{equation}
                2A b^{\prime\prime} r^2 - 2A bb^{\prime\prime}r
                + 4r A b^\prime - 4Abb^\prime +kr{b^\prime}^2 -
                b b^\prime = 0
                \label{Eq11}
          \end{equation}
where $ \frac{k}{1-k} = A $ . \\

Now to investigate whether there exists physically meaningful
solutions consistent with the boundary requirements [ conditions
(1) to (4) ], we take a general functional form of $b(r)$ . We
can generally express it in the form
\begin{equation}
                b(r) = \Sigma_{n=1}^{\infty} b_n r^n +
                \Sigma_{m=0}^{\infty} a_m r^{-m}
                \label{Eq12}
          \end{equation}
since $\frac{b(r)}{r} \rightarrow 0 $ as $ r \rightarrow \infty
$, equation (12) is consistent only when all the $ b_n$'s in
$b(r)$ vanish i.e.
\begin{equation}
                b(r) =  \Sigma_{m=0}^{\infty} \frac{a_m}{r^m}
                \label{Eq13}
          \end{equation}
Plugging this in equation (11) and matching the co-efficients of
equal powers of $r$ from both sides , we get ,\\
for $ k \neq 0 $
\begin{eqnarray*} \\&&
                b(r) =  a_0 + \frac{a_1}{r}-\frac{a_0a_1}{4Ar^2}+\frac{1}{12Ar^3}
                [a_0^2a_1(\frac{1}{2A}-1)-(k+1)a_1^2]+\\&&\frac{1}{24Ar^4}
                [a_0^3a_1(\frac{1}{2A}-1)(1-\frac{1}{4A})-a_0a_1^2( (k+2)- \frac{4+5k}{4A})]
                + .................
                \label{Eq14}
         \end{eqnarray*}
Thus we get two parameters family of solutions . \\

Now the expression for $\rho$ can be obtained from equation (8) as
\begin{equation}
                \rho =
                \frac{1}{8\pi}[-\frac{a_1}{r^4}+\frac{a_0a_1}{2Ar^5}-
                \frac{1}{4Ar^6}[a_0^2a_1(\frac{1}{2A}-1)-(k+1)a_1^2]
                +................]
            \label{Eq15}
          \end{equation}

Since k is either $ >1$ or $<-1$, so A lies in $(-1,-\frac{1}{2}
)$. Thus $ \rho>0$ implies $a_1$ should be negative.

 The equation (10) gives the following expression for $'f'$
as
\begin{equation}
                e^{2f} = [ \frac{\rho_0}{8\pi[-a_1+\frac{a_0a_1}{2Ar}-
                \frac{1}{4Ar^2}[a_0^2a_1(\frac{1}{2A}-1)-(k+1)a_1^2]
                +................]}]^{\frac{2k}{k-1}}
            \label{Eq16}
          \end{equation}
The throat of the wormhole occurs at $ r=r_0$ where $r_0$ is the
solution of the equation $ b(r) = r$. Suppose $\frac{1}{r} = y$,
then $ b(r) = r$ implies
\begin{equation} g(y) = a_0y + a_1y^2 -
                \frac{a_0a_1}{4A} y^3+ ............ - 1 = 0
            \label{Eq17}  \end{equation}
This is a polynomial equation with negative last term. Then this
equation must have at least one positive root, say, $ y =
\frac{1}{r_0} $. Since $\frac{1}{r_0} $ is a root of equation
(16), then by standard theorem of algebra, either $g(y) > 0$ for $
y > \frac{1}{r_0}$ and $g(y) < 0 $ for $ y < \frac{1}{r_0}$ or
$g(y) < 0$ for $ y > \frac{1}{r_0}$ and $g(y) > 0$ for $ y <
\frac{1}{r_0}$. Let us take the first possibility and one can
note that for $ y = \frac{1}{r} < \frac{1}{r_0}$ i.e. $r>r_0$,
$g(y) < 0$, in other words,  $b(r) < r $. But when $ y =
\frac{1}{r} > \frac{1}{r_0}$ i.e. $r<r_0$, $g(y) > 0$, this means,
$b(r)
> r $, which violates the wormhole structure given in equation(1).

One can note that the redshift function $ f(r)$ always finite for
$ r \geq r_0 > 0 $ i.e. no horizon exists in the space time .
Thus our solution describing a static spherically symmetric
wormhole supported by the Phantom Energy . \\

The asymptotical wormhole mass reads
\begin{equation}
                M = \lim_{r \rightarrow \infty} \frac {1}{2} b (r) = \frac{a_0}{2}
            \label{Eq18}
          \end{equation}
The axially symmetric embedded surface $ z = z(r)$ shaping the
wormhole's spatial geometry is a solution of
\begin{equation}
                \frac{dz}{dr}=\pm
                \frac{1}{\sqrt{\frac{r}{b(r)}-1}}
            \label{Eq19}
          \end{equation}
By the definition of wormhole , we can note that at the value $ r
= r_0 $ (the wormhole throat) equation (18) is divergent i.e.
embedded surface is vertical there .\\
\pagebreak

 According to Morris and Throne [3] , the 'r'
co-ordinate is
ill-behaved near the throat, but proper radial distance\\
\begin{equation}
 l(r) = \pm \int_{r_0^+}^r \frac{dr}{\sqrt{1-\frac{b(r)}{r}}}
            \label{Eq20}
          \end{equation}
 must be well behaved everywhere i.e. we must require that $ l(r)
 $is finite throughout the space-time . \\

 For our Model,
\begin{equation}
 l(r) = \pm \int_{r_0^+}^r
 \frac{dr}{\sqrt{1-\frac{1}{r}[a_0+\frac{a_1}{r}-\frac{a_0a_1}{4Ar^2}+....]}}
            \label{Eq21}
          \end{equation}
Though we can not find the explicit form of the integral but one
can see that the above integral is a convergent integral i.e.
proper length should be finite . \\

To summarize , we have constructed exact solution describing
static symmetric wormholes supported by the Phantom Energy . The
resulting line element represents a two parameter family of
geometries which contains wormholes . Hence this shows clearly
that the Phantom Energy can support the existence of static
wormholes .

For $a_1 = 0 $, we obtain the standard Schwarzschild solution,
viz., $ e^{2f(r)} = [1 - \frac{b(r)}{r}] = 1 - \frac{r_s}{r}$,
provided $ a_0 = r_s = 2GM $ (  the Schwarzschild radius).

  One of the most striking feature of our model
is that if we choose k is very close to unity, then  $ p + \rho
 = ( - k +1 )\rho =  \frac{( - k +1 )}{8\pi}[-\frac{a_1}{r^4}+\frac{a_0a_1}{2Ar^5}-
                \frac{1}{4Ar^6}[a_0^2a_1(\frac{1}{2A}-1)-(k+1)a_1^2]
                +..............]$ can be made arbitrarily small.
                This reveals the fact that there is  possible
                to
                construct a wormhole by an arbitrarily small
                amount of Phantom Energy.

 Finally , if we
take the parameter $a_0 > 0 $, then asymptotic mass $'M'$ of the
Phantom Energy wormhole is positive i.e. a distant observer could
not see any difference of gravitational nature between wormhole
and a compact mass 'M'.

        { \bf Acknowledgements }

          F.R is thankful to Jadavpur University and DST , Government of India for providing
          financial support under Potential Excellence and Young
          Scientist scheme . We are grateful to anonymous referee for pointing out the errors
          of the paper and for his
          constructive suggestions, which has led to a stronger result than the one in the
          original version.  \\
\pagebreak


\end{document}